# Metamaterials Demonstrating Negative Thermal Capacity

**Edward Bormashenko *[1], Irina Legchenkova[1], Evgeny Shulzinger**

[1]Department of Chemical Engineering, Engineering Faculty, Ariel University, Ariel, Israel, 407000
*Correspondance: edward@ariel.ac.il (E.B.)

**Abstract:** One-dimensional chain of core-shell pairs connected by ideal springs enables design of the metamaterial demonstrating the negative effective density and negative specific thermal capacity. We assume that the molar thermal capacity of the reported metamaterial is governed by the Dulong-Petit law in its high temperature limit. The specific thermal capacity depends on the density of the metamaterial; thus, it is expected to be negative, when the effective density of the chain is negative. The range of the frequencies enabling the effect of the negative thermal capacity is established. Dependence of the effective thermal capacity on the exciting frequency for various core/shell mass ratios is elucidated. The effective thermal capacity becomes negative in the vicinity of the local resonance frequency $\omega_0$ in the situation when the frequency $\omega$ approaches $\omega_0$ from above. The effect of the negative effective thermal capacity is expected in metals in the vicinity of the plasma frequency. The effect of the viscous dissipation on the effective thermal capacity is addressed.

## 1. Introduction

21-st Century will demand for new materials and for the novel ideology of materials science, considering rapid depletion of natural resources and sustainability of the human society. Among the novel trends in the modern materials science the development of meta-materials is prominent [1-7]. Metamaterials are recently developed artificial materials demonstrating properties that are not found in naturally occurring materials. The domain of metamaterials covers a broad diversity of fields in physics and engineering, including electromagnetics [3,8-9], acoustics [9-13], mechanics [5, 7] and thermodynamics [14]. In particular, metamaterials demonstrate negative values of refractive index [1-3, 8-9, 11], bulk modulus [11, 15], Poisson Ratio [16, 17] and thermal expansion [18-20].

Our paper is devoted to the design of the metamaterial in which specific thermal capacity is expected to be negative. This effect becomes possible due to development of the metamaterials possessing a negative effective density [21-29]. It was demonstrated that the dynamic effective mass (density) of an inhomogeneous mixture, used in the prediction of wave velocities in the long wavelength limit, can differ from the static version, namely the volume average of the component mass densities [21, 23, 27, 29]. The effective mass density of certain kinds of elastic media built of mass-spring-shell oscillators becomes frequency dependent and may become negative for frequencies near the resonance frequency [23, 29]. In our recent works we demonstrated that the effect of negative density may be achieved in the vicinity of frequency of plasma oscillations of electron gas embedded within the crystalline lattice [30-31]. Now, we demonstrate that the effect of negative density yields in a somewhat paradoxical way the negative effective thermal capacity solids.

## 2. Results and discussion
### 2.1. Negative thermal capacity in the condensed matter demonstrating negative effective density

Consider the 1D chain of oscillators presented in **Figure 1** and treated in refs. [23, 28-31]. The 1D lattice, shown in **Figure 1** is built from the core/shell oscillator pairs connected by elastic springs. A core mass $m_2$ is connected through the ideal spring with constant $k_2$ to a rigid shell with mass $m_1$. The system is excited by the harmonic force $F = \hat{F} sin \omega t$. If we solve the Newtonian equations of motion for the masses $m_1$ and $m_2$ and replace the core/shell oscillator with a single effective mass $m_{eff}$, we derive Equation 1:

$$m_{eff} = m_1 + \frac{m_2 \omega_0^2}{\omega_0^2 - \omega^2} \qquad (1)$$

where $\omega_0 = \sqrt{\frac{k_2}{m_2}}$. It is immediately seen from Eq. 1 that when the frequency $\omega$ approaches $\omega_0$ from above the effective mass $m_{eff}$ will be negative [23, 28-31]. The extension of the aforementioned approach to continuous medium leads to the idea of the effective negative density [28-29]. The effective density of the chain depicted in **Figure 1** $\rho_{eff}(\omega)$, was calculated in ref. 29; and it is given by Eq. 2:

$$\rho_{eff}(\omega) = \rho_{st} \frac{\theta}{\delta(1+\theta)\left(\frac{\omega}{\omega_0}\right)^2} \left\{ cos^{-1} \left\{ 1 - \frac{\delta}{2\theta} \frac{\left(\frac{\omega}{\omega_0}\right)^2 \left[ \left(\frac{\omega}{\omega_0}\right)^2 - (1+\theta) \right]}{\left(\frac{\omega}{\omega_0}\right)^2 - 1} \right\} \right\}^2, \qquad (2)$$

where the static linear density of the chain $\rho_{st}$ is given by: $\rho_{st} = \frac{m_1 + m_2}{a}$; $[\rho_{st}] = \frac{kg}{m}$ and $\theta = \frac{m_2}{m_1}$; $\delta = \frac{k_2}{k_1}$; and $a$ is the lattice constant (see **Figure 1**). The realization of the elastic medium with the negative effective density exploiting strong Mie-type resonance was suggested in ref. 28. Experimental realization of the elastic medium with the negative effective density was reported in ref. 32.

We demonstrate that the effective negative mass (density) media also give rise to the effect of negative effective thermal capacity. Consider 1D lattice of the core/shell pairs connected with ideal springs $k_1$, shown in **Figure 1**. Let us estimate the effective thermal capacity of this 1D lattice. In the limit of high temperatures the molar thermal capacity of the aforementioned chain (lattice) will be described by the modified Dulong-Petit equation [33-35]:

$$c_{mol} = const \cong R \ , \qquad (3)$$

where $c_{mol}$ is the thermal capacity of the 1D ensemble built of Avogadro number of the core/shell oscillators, depicted in **Figure 1**, as established under the constant volume. The multiplier "3" is omitted in Eq. 3 due to the fact that we deal with the 1D chain of oscillators. It is noteworthy that the Dulong-Petit law works well in the realms of the both of classical and quantum mechanics, if the springs connecting the elements of lattice are supposed to be ideal [33-37]. The deviations from the Dulong-Petit law become essential when anharmonic effects are considered [38]; we restrict our treatment by the assumption that elastic springs are ideal. The effective dynamic specific thermal capacity $c_{eff}(\omega)$ of the 1D medium, built of the oscillators, shown in **Figure 1** is estimated as:

$$c_{eff}(\omega) \cong \frac{R}{\rho_{eff}(\omega) V_{mol}}, \qquad (4)$$

where $R$ is the gas constant, $V_{mol}$ is the linear molar volume; $[V_{mol}] = m$; $[\rho_{eff}] = \frac{kg}{m}$ and the dimension of the specific thermal capacity is: $[c_{eff}] = \frac{J}{kg \times K}$. Combining Equations (2) and (4) immediately yields:

$$c_{eff}(\omega) = \frac{\delta(1+\theta)\left(\frac{\omega}{\omega_0}\right)^2 R}{\rho_{st} V_{mol} \theta \left\{ cos^{-1} \left\{ 1 - \frac{\delta}{2\theta} \frac{\left(\frac{\omega}{\omega_0}\right)^2 \left[ \left(\frac{\omega}{\omega_0}\right)^2 - (1+\theta) \right]}{\left(\frac{\omega}{\omega_0}\right)^2 - 1} \right\} \right\}^2} \ . \qquad (5)$$

Let us denote: $c_0 = \frac{R\delta(1+\theta)}{\rho_{st}v_{mol}\theta} > 0$ and $\tilde{C}_{eff}(\omega) = \frac{c_{eff}(\omega)}{c_0}$, which is the dimensionless effective thermal capacity.

Thus Eq. 5 appears in its dimensionless form as follows:

$$\tilde{C}_{eff}(\omega) = \frac{\left(\frac{\omega}{\omega_0}\right)^2}{\left\{cos^{-1}\left\{1-\frac{\delta}{2\theta}\frac{\left(\frac{\omega}{\omega_0}\right)^2\left[\left(\frac{\omega}{\omega_0}\right)^2-(1+\theta)\right]}{\left(\frac{\omega}{\omega_0}\right)^2-1}\right\}\right\}^2} \quad . \tag{6}$$

Obviously, the field of frequencies in which $\rho_{eff}(\omega) < 0$ takes place implies $\tilde{C}_{eff}(\omega) < 0$. It is convenient to introduce the following designations: $\beta = \frac{\delta}{2\theta} > 0$; $\Omega = \frac{\omega}{\omega_0} > 0$. Thus, Eq. 5 may be re-written as follows:

$$\tilde{C}_{eff}(\Omega) = \frac{\Omega^2}{\left\{cos^{-1}\left\{1-\beta\frac{\Omega^2[\Omega^2-(1+\theta)]}{\Omega^2-1}\right\}\right\}^2}. \tag{7}$$

The effective thermal capacity $\tilde{C}_{eff}(\Omega)$ will be complex, when Eq. 8 takes place:

$$\frac{\Omega^2[\Omega^2-(1+\theta)]}{\Omega^2-1} < 0. \tag{8}$$

Eq. 8 immediately yields for $\Omega$, at which inequality (8) takes place:

$$0 < \Omega^2 - 1 < \theta \ , \tag{9}$$

which is defining the range of frequencies at which the complex and in particular negative effective thermal capacity is expected. The dependencies $\tilde{C}_{eff}(\Omega)$ are depicted in **Figures 2-5**. The MATLAB software was used for the preparing of the plots. We fixed the values of $c_0$ and $\beta$, namely: $c_0 = 1\frac{J}{kg\times K}$, $\beta = 1/2$, varied $\theta$ in the range $10^{-3} < \theta < 10^3$ and plotted $\tilde{C}_{eff}(\Omega)$ in **Figures 2-5**.

Various configurations of the positive and negative branches of $\tilde{C}_{eff}(\Omega)$ are depicted in **Figures 2-5**. Consider, that generally the effective thermal capacity $\tilde{C}_{eff}(\Omega)$ is given by a complex number. We restrict our treatment within the range of frequencies supplying real values of effective thermal capacity (positive or negative), as shown in **Figures 2-5**. It should be emphasized that the critical frequency $\Omega_{cr}$ separating areas of positive and negative thermal capacities depends on the interrelation of masses $= \frac{m_2}{m_1}$ , as it recognized from Eq. 8 and **Figures 2-5**. The analytical expression for $\Omega_{cr}$ emerging from Eq. 7 is supplied by Eq. 10:

$$\Omega_{cr} = \sqrt{1+\theta} \ . \tag{10}$$

When the core mass $m_2$ is markedly smaller than the shell mass $m_1$ (i.e. $\theta \ll 1$) the effective thermal capacity becomes negative in the vicinity of the local resonance frequency $\omega_0$, in the situation when the frequency $\omega$ approaches $\omega_0$ from above ($\Omega_{cr} \cong 1+\frac{\theta}{2}$), as shown in **Figure 2**. The case of the "light core mass" corresponds to the plasma oscillations of the free electron gas in metals, as discussed in detail in refs. 30-31 and depicted schematically in **Figure 6**. Thus, the effect of the negative effective thermal capacity is expected in metals in the vicinity of the plasma frequency.

It is instructive to establish the asymptotic behavior of Eq. 7, when $\theta \to 0$; $\beta\Omega^2 \to 0$ takes place. The asymptotic is supplied by Eq. 11:

$$\lim_{\theta \to 0; \beta \Omega^2 \to 0} \tilde{C}_{eff}(\Omega) = \frac{\Omega^2}{cos^{-1}(1-\beta\Omega^2)} \cong \sqrt{\frac{2}{\beta}}\Omega \ . \tag{11}$$

It is recognized from **Figures 2-5** that the effective thermal capacity $\tilde{C}_{eff}(\Omega)$ demonstrates a behavior typical for the first order phase transitions, namely it exhibits discontinuity at $\Omega = \Omega_{cr}$ [33-35, 39]. However, this kind of the phase transition, may be called "pseudo-phase-transition" due the fact that the molar thermal capacity of the discussed lattice remains constant, and it is dictated by the Dulong-Petit rule.

What is the physical meaning of the "negative effective thermal capacity"? It seems that it is absurd, which contradicts to the energy conservation. The close inspection of the problem shows that the addressed core/shell lattice demonstrates the negative effective thermal capacity when exerted to the external harmonic vibrations; in other words, it is excited by the external source of energy; thus, there is no conflict with the energy conservation. The more subtle problem is the analysis of the Second Law of Thermodynamics as applied for the addressed system, namely establishment of the direction of the energy flow for the chain built of core-shell oscillators in the vicinity of the internal resonance frequency. We plan to address this problem in our future investigations.

## 2.2. Considering friction in the 1D chain of oscillators possessing negative effective thermal capacity

Now consider the effect of friction on the effect of the negative thermal capacity. Let springs $k_2$ in **Figure 1** be viscoelastic ones. If springs $k_2$ behaves according to a Maxwell model of a dashpot in series with a purely elastic spring, then:

$$\frac{1}{k_2} = \frac{1}{K_2} + \frac{i}{\omega\eta} \ , \tag{12}$$

where $i = \sqrt{-1}$, $K_2$ and $\eta$ are the real modulus of the purely elastic spring component and the coefficient of viscosity of the dashpot component, respectively. Consequently, the effective mass appears as (see ref. 23):

$$m_{eff} = m_1 + \frac{m_2\omega_0^2}{\omega_0^2 - \omega^2 - \frac{i\omega}{\eta}} \ , \tag{13}$$

where $\omega_0 = \sqrt{\frac{K_2}{m_2}}$. Simple transformations yield:

$$\frac{1}{k_2} = \frac{1}{m_2\omega_0}(\Omega + iD), \tag{14}$$

where the dimensionless real and imaginary components of the frequency are defined as follows: $\Omega = \frac{\omega}{\omega_0}; D = \frac{m_2\omega_0}{\eta} = \frac{\sqrt{K_2 m_2}}{\eta}$. Thus, the effective thermal capacity $\tilde{C}_{eff}(\Omega')$ may be seen as a function of the dimensionless complex frequency $\Omega' = \Omega + iD; 0 < D < 1$. The critical frequency is now supplied by Eq. 15:

$$\Omega_{cr} = \sqrt{(1 + \theta - D^2)}. \tag{15}$$

Amplitude of the vibrations $A$ now is fading with time, and it scales as $A \sim e^{-D\omega_0 t}$. Real and imaginary components of the effective thermal capacity dependence $\tilde{C}_{eff}(\Omega')$ calculated numerically for the different values of dimensionless parameters $\theta = \frac{m_2}{m_1}$ and $D$ are supplied in **Figures 7-8**.

## 3. Conclusions

We report one-dimensional metamaterials demonstrating the effective negative thermal capacity within the certain range of frequencies of the exciting force acting on the material. The metamaterial is built from the 1D chains comprising core/shell oscillators connected with the ideal springs $\omega_0$. It was demonstrated that the effective density of such a chain if frequency-dependent and it may be negative, when the frequency of the external force approaches the resonance frequency from above [23, 28-31]. We adopt that the Dulong-Petit law is valid in the high temperature limit for the molar thermal capacity of the introduced metamaterial [33-37]. The specific effective thermal capacity $[c_{eff}] = \frac{J}{kg \times K}$ depends on the effective dynamic density of the chain; and, thus, it depends on the frequency of the exciting force $\omega$.

The dimensionless $\omega$-dependent effective thermal capacity is defined as $\tilde{C}_{eff}(\omega) = \frac{c_{eff}(\omega)}{c_0}$, where $c_0 = \frac{R\delta(1+\theta)}{\rho_{st}V_{mol}\theta}$, where $\theta = \frac{m_2}{m_1}$; $\delta = \frac{k_2}{k_1}$. It is instructive to introduce the dimensionless frequency $\Omega = \frac{\omega}{\omega_0}$ and to calculate the dependence $\tilde{C}_{eff}(\Omega)$. The value of $\tilde{C}_{eff}(\Omega)$ turns out to be complex for the certain range of dimensionless frequencies $\Omega = \frac{\omega}{\omega_0}$. The frequency dependence of the dimensionless effective thermal capacity $\tilde{C}_{eff}(\Omega)$ is influenced by the core/shell mass ratio $\theta = \frac{m_2}{m_1}$. The critical frequency $\Omega_{cr}$ separating areas of positive and negative thermal capacities is established as $\Omega_{cr} = \sqrt{1+\theta}$. The dimensionless effective thermal capacity $\tilde{C}_{eff}(\Omega)$ demonstrates a behavior typical for the first order phase transitions, namely it exhibits discontinuity at $\Omega = \Omega_{cr}$ (recall that the molar thermal capacity of the metamaterial remains constant and it is governed by the Dulong-Petit law). The effect of the negative effective thermal capacity is predicted for metals in the vicinity of the electron gas plasma oscillations frequency, corresponding to: $\theta \ll 1$; $\Omega_{cr} \cong 1 + \frac{\theta}{2}$. The effect of the "negative effective thermal capacity" does not conflict with the energy conservation. The investigated core/shell lattice demonstrates the negative effective thermal capacity when exerted to the external harmonic vibrations; in other words, it is excited by the external source of energy; so, there is no contradiction to the energy conservation law. The effect of the viscous dissipation on the effective thermal capacity is treated numerically.

**Acknowledgements:** The authors are indebted to Mrs. Yelena Bormashenko for her kind help in preparing this work.

**Author Contributions**: Conceptualization, E.B., E.S.; methodology, E.B., E.S., I.L.; software, E.S., I.L.; validation, E. B., E.S., I.L.; formal analysis E. B., E.S.; investigation, E. B., E.S., E.B.; I.L.; writing—original draft preparation, E. B., I.L.; writing—review and editing; E.B., E.S., I.L.; supervision, E. B.; project administration, E.B.; funding acquisition, E.B. All authors have read and agreed to the submitted version of the manuscript.

**Funding:** No external funding was obtained for this work.

**Conflicts of Interest:** The authors declare no conflict of interests.

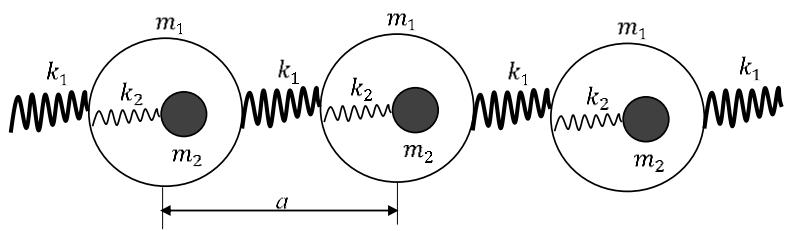

Figure 1. The chain of core/shell pairs connected by ideal springs $k_1$ make possible 1D medium possessing negative effective density and thermal capacity.

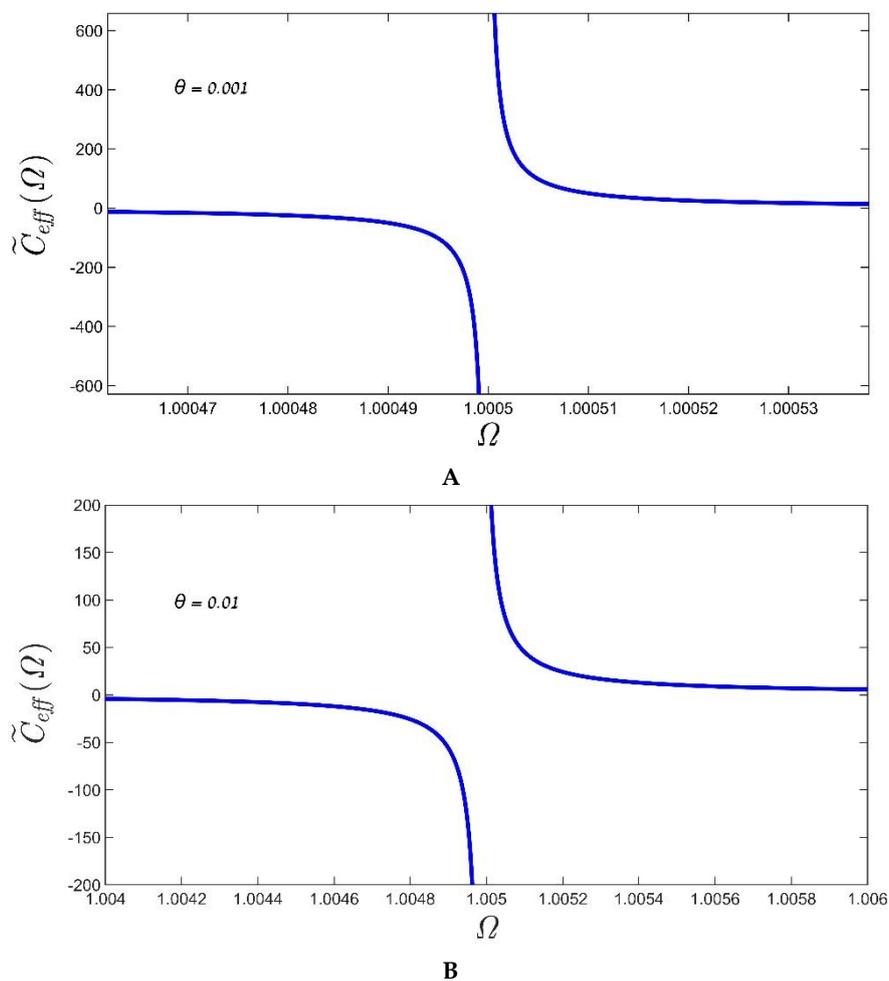

Figure 2. The dependence $\widetilde{C}_{eff}(\Omega); \Omega = \frac{\omega}{\omega_0}$ is depicted; A. $\theta = 0.001$; B. $\theta = 0.01$.

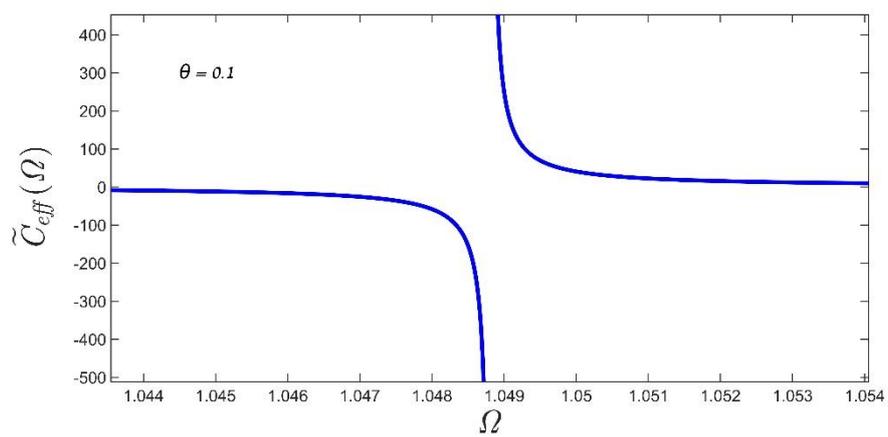

**A**

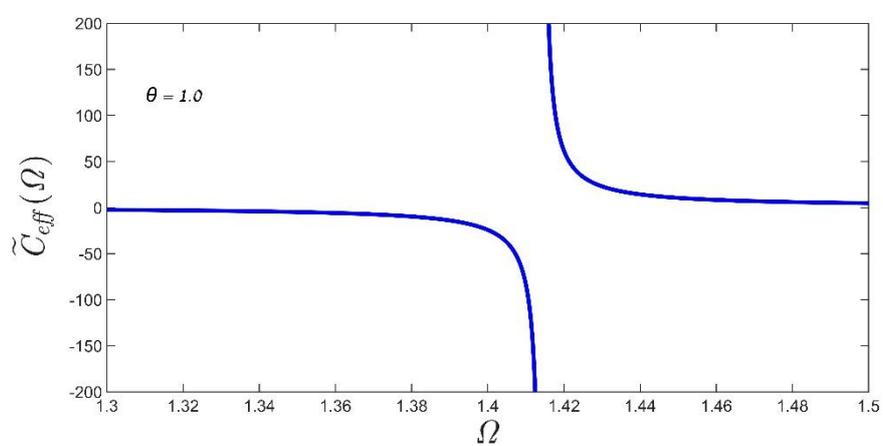

**B**

Figure 3. The dependence $\widetilde{C}_{eff}(\Omega); \Omega = \frac{\omega}{\omega_0}$ is depicted; **A. $\theta = 0.1$; B. $\theta = 1$.**

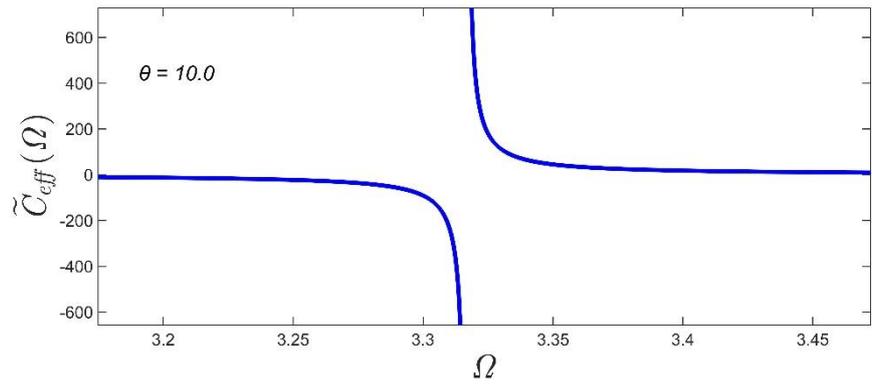

**A**

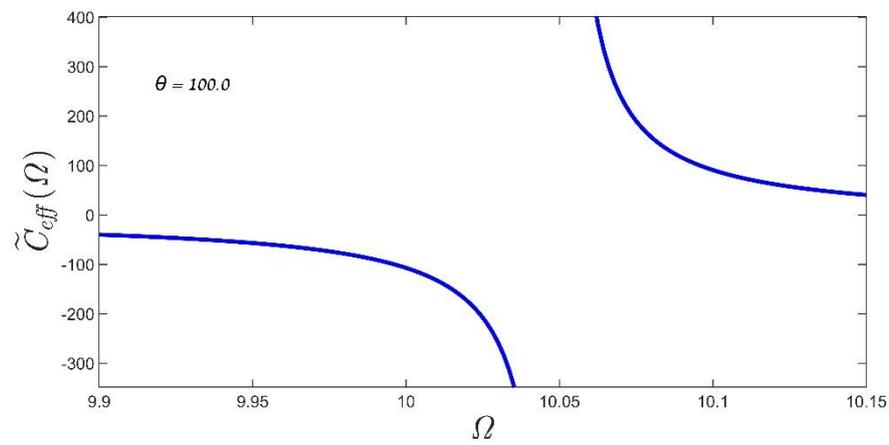

**B**

Figure 4. The dependence $\widetilde{C}_{eff}(\Omega)$; $\Omega = \frac{\omega}{\omega_0}$ is depicted; **A**. $\boldsymbol{\theta = 10}$; **B**. $\boldsymbol{\theta = 100}$.

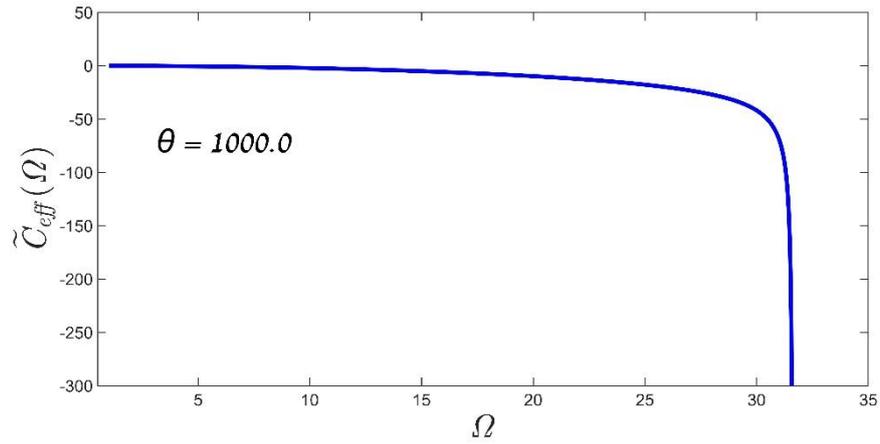

Figure 5. The dependence $\widetilde{C}_{eff}(\Omega); \Omega = \frac{\omega}{\omega_0}$ is depicted; $\theta = 1000$.

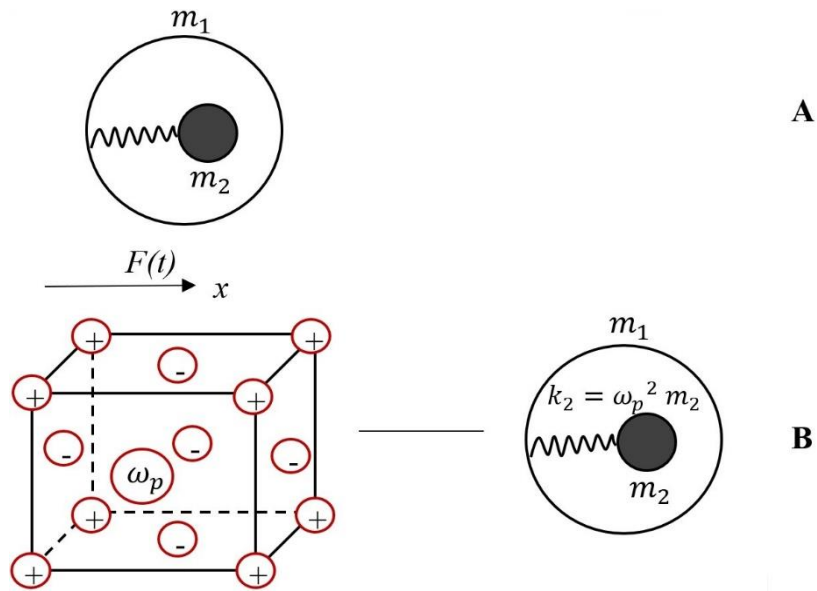

Figure 6. **A.** Core with mass $m_2$ is connected internally through the spring with $k_2$ to a shell with mass $m_1$. The system is subjected to the sinusoidal force $F(t) = \widehat{F}sin\omega t$. **B.** Free electrons gas $m_2$ is embedded into the ionic lattice $m_1$; $\omega_p$ is the plasma frequency (the left sketch). The equivalent mechanical scheme of the system (right sketch).

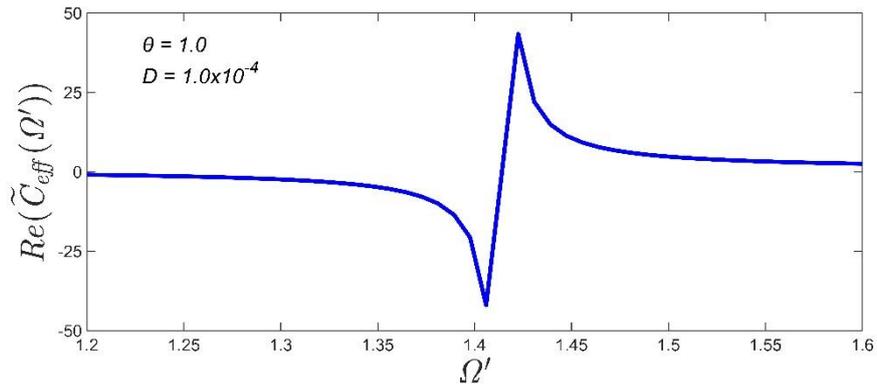

**A**

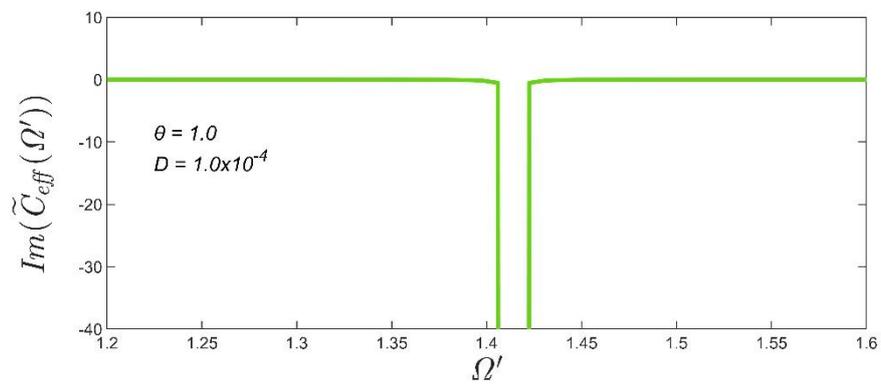

**B**

Figure 7. Considering of the effect of friction on the thermal capacity is illustrated for $\boldsymbol{\theta = 1.0}; \boldsymbol{D = 1.0 \times 10^{-4}}$. Real (**A**) and imaginary (**B**) parts of the numerically calculated effective thermal capacity $\widetilde{\mathcal{C}}_{eff}(\Omega')$ are presented. $\boldsymbol{\Omega' = \Omega + iD}$. (See Section 2.2).

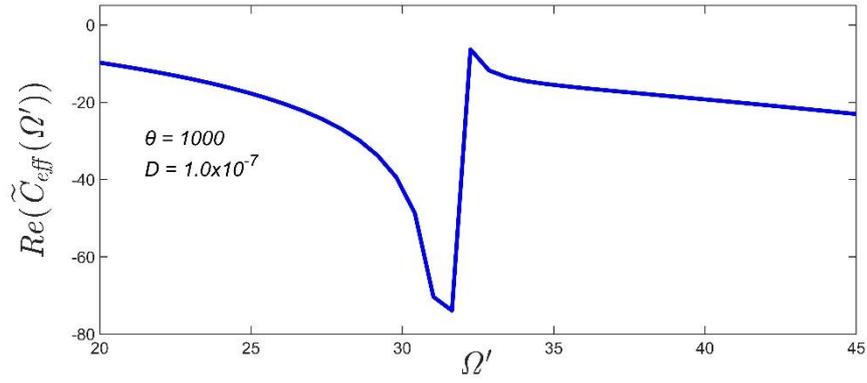

**A**

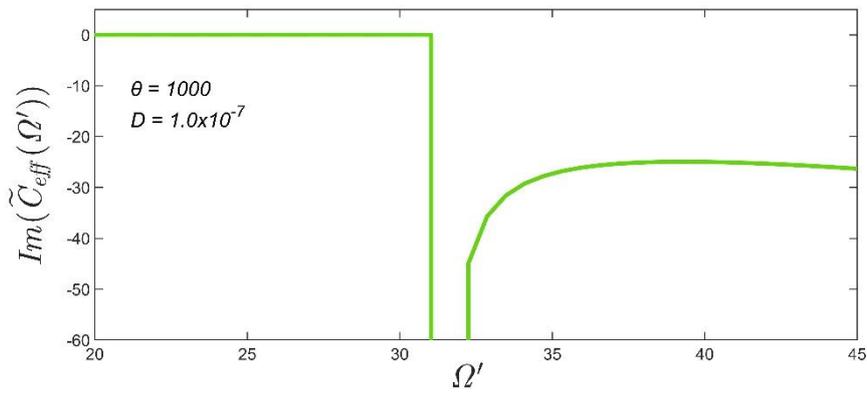

**B**

Figure 8. Considering of the effect of friction on the thermal capacity is illustrated for $\boldsymbol{\theta = 10^3; D = 1.0 \times 10^{-7}}$. Real (**A**) and imaginary (**B**) parts of the numerically calculated $\widetilde{\mathcal{C}}_{eff}(\Omega')$ thermal capacity are presented. $\boldsymbol{\Omega' = \Omega + iD}$ (see Section 2.2).

Table of Contents Image

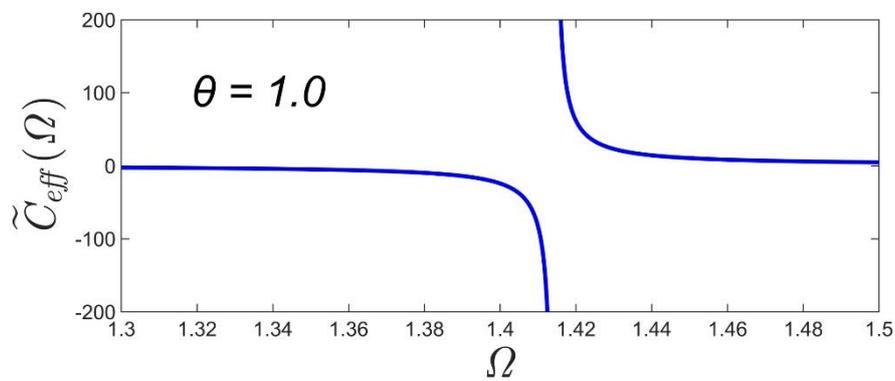